\documentclass[amsmath,amssymb,aps,prd,11pt,tightenlines,superscriptaddress,nofootinbib,preprintnumbers,notitlepage]{revtex4-1}

\usepackage{graphicx}
\usepackage{dcolumn}
\usepackage{bm}
\usepackage{xcolor}
\usepackage{xspace}
\usepackage{subfigure}
\usepackage{topcapt}
\usepackage{lineno}

\newcommand{\mCP}{\ensuremath{\mathrm{MCP}}\xspace}
\newcommand{\mCPp}{\ensuremath{\mCP^+}\xspace}
\newcommand{\mCPm}{\ensuremath{\mCP^-}\xspace}

\begin{document}

\title{Searching for millicharged particles with the FORMOSA experiment at the CERN LHC}
\author{Matthew Citron}\affiliation{University of California, Davis,  California 93106, USA}
\author{Frank Golf}\affiliation{Boston University, Boston,  Massachusetts 02215, USA}
\author{Kranti Gunthoti}\affiliation{Los Alamos National Laboratory (LANL), Los Alamos, NM 87545, USA}
\author{Andrew~Haas}\affiliation{New York University, New York, New York 10012, USA}
\author{Christopher~S.~Hill}\affiliation{The Ohio State University, Columbus, Ohio 43218, USA}
\author{Dariush~Imani}\affiliation{University of California, Santa Barbara,  California 93106, USA}
\author{Samantha~Kelly}\affiliation{University of California, Davis,  California 93106, USA}
\author{Ming~Liu}\affiliation{Los Alamos National Laboratory (LANL), Los Alamos, NM 87545, USA}
\author{Steven~Lowette}\affiliation{Vrije Universiteit Brussel, Brussel 1050, Belgium}
\author{Albert~De~Roeck}\affiliation{CERN, Geneva 1211 Switzerland}
\author{Sai~Neha~Santpur}\affiliation{University of California, Santa Barbara,  California 93106, USA}
\author{Ryan~Schmitz}\affiliation{University of California, Santa Barbara,  California 93106, USA}
\author{Jacob~Steenis}\affiliation{University of California, Davis,  California 93106, USA}
\author{David~Stuart}\affiliation{University of California, Santa Barbara,  California 93106, USA}
\author{Yu-Dai~Tsai}\affiliation{Los Alamos National Laboratory (LANL), Los Alamos, NM 87545, USA}
\affiliation{University of California, Irvine,  California 92617, USA}
\author{Juan~Salvador~Tafoya~Vargas}\affiliation{University of California, Davis,  California 93106, USA}
\author{Tiepolo~Wybouw}\affiliation{Vrije Universiteit Brussel, Brussel 1050, Belgium}
\author{Jaehyeok~Yoo}\affiliation{Korea University, Seoul, South Korea.}

\begin{abstract}
\noindent In this contribution, we evaluate the sensitivity for particles with charges much smaller than the electron charge with a dedicated scintillator-based detector in the far forward region at the CERN LHC, FORMOSA. This contribution will outline the scientific case for this detector, its design and potential locations, and the sensitivity that can be achieved. The ongoing efforts to prove the feasibility of the detector with the FORMOSA demonstrator will be discussed. Finally, possible upgrades to the detector through the use of high-performance scintillator will be discussed.

\end{abstract}
\maketitle
\newpage

\section{Scientific context}
Over a quarter of the mass-energy of the Universe is widely thought to be some kind of nonluminous ``dark'' matter (DM); however, all experiments to date have failed to confirm its existence as a particle, much less measure its properties. The possibility that DM is not a single particle, but rather a diverse set of particles with as complex a structure in their sector as normal matter, has grown in prominence in the past decade, beginning with attempts to explain observations in high-energy astrophysics experiments~\cite{ArkaniHamed:2008qn, Pospelov:2008jd}.   
Many experimental efforts have been launched to look for signs of a dark sector, including searches at high-energy colliders, explorations at low-energy colliders, precision tests, and effects in DM direct detection experiments (for recent reviews see Refs.~\cite{Battaglieri:2017aum, Beacham:2019nyx,Strategy:2019vxc}). 
Most of these experiments target the dark sector via massive mediators, such as a massive dark photon. However, in theories in which the dark photon is massless, the principal physical effect is that new dark sector particles that couple to the dark photon will have a small effective electric charge~\cite{Holdom:1985ag, Izaguirre:2015eya}. While direct searches robustly constrain the parameter space of millicharged particles (MCPs), indirect observations can be evaded by adding extra degrees of freedom, which can readily occur in minimally extended dark sector models~\cite{Izaguirre:2015eya}. In particular, the parameter space 1 $ < m_\mCP < 100$~GeV is largely unexplored by direct searches~\cite{MilliQ,essig2013dark,Chatrchyan_2013,Chatrchyan_2013_2,Acciarri_2020, Davidson:2000hf, Badertscher:2006fm,Magill_2019}.

The high energy of the LHC beams can efficiently produce millicharged particles with masses $m_\mCP$ up to $\sim45$~GeV. The milliQan experiment~\cite{Ball:2016zrp} has been running in the CMS service cavern at the LHC since 2018. The milliQan demonstrator provided the first sensitivity to MCPs at a hadron collider using $\sqrt{s} = 13$ TeV data collected during LHC Run 2~\cite{Ball_2020} and an upgraded detector will extend this sensitivity by another order of magnitude with $\sqrt{s} = 13.6$ TeV data collected during LHC Run 3~\cite{Ball_2021}. 
The mass range for which the LHC provides unique sensitivity ($\gtrsim$ GeV) is small compared to the LHC center of mass energy of 13.6 TeV. Given this small mass compared to transverse momentum, these particles will be produced with an approximately flat distribution in pseudorapidity ($\eta$). Therefore, a millicharged particle detector placed at $\eta \sim 7$ would be expected to see around factor 250 higher rate of MCPs compared to a detector at $\eta \sim 0$ as the forward detector intersects a much larger range in $\eta$ as well as having full coverage in $\phi$. 

This contribution will outline the progress towards designing, constructing and operating a forward millicharged particle detector, named FORMOSA ~\cite{Tsai:2020ForMINI} to extend the search for MCPs by multiple orders of magnitude in mass and charge. This detector can be placed in the Forward Physics Facility (FPF)~\cite{Feng_2023}, or, less optimally, in an existing location at the LHC. The FORMOSA experiment provides an excellent opportunity to study models of DM in which the dark sector communicates with the standard model via a massless dark photon. The FORMOSA experiment can also provide sensitivity to other dark sector phenomena such as ``quirks" produced in the forward region using a strategy similar to that outlined in Ref.~\cite{Feng_2024}. The FORMOSA experiment will help to ensure that CERN fulfills  the recommendations from past EPPSU recommendations to diversify its physics program through the construction and operation of a wide range of smaller detectors that provide unique sensitivity beyond that which can be achieved by the main experiments of the laboratory’s collider program~\cite{Strategy:2019vxc}. Indeed, the FORMOSA experiment was explicitly highlighted in the US Particle Physics Project Prioritization Panel report as a clear example of an ``agile'' experiment that will provide unique new sensitivity to the dark sector~\cite{Asai:2905898}. The FORMOSA experiment receives support from the CERN Physics Beyond Colliders Study group, which has submitted a separate contribution that summarizes sensitivities to a wide range of new phenomena achieved by dedicated experiments proposed at CERN.  

In this contribution, the experimental design is discussed in Sec~\ref{sec:experimentalDesign}, the sensitivity to millicharged particles is detailed in Sec~\ref{sec:sensitivity}, the operation of a prototype detector is outlined in Sec~\ref{sec:proto}, the progress towards constructing the experiment is detailed in Sec~\ref{sec:planning}, and finally, a summary is provided in Sec~\ref{sec:summary}.

\section{Experimental design}
\label{sec:experimentalDesign}

\begin{figure}[tbp]
  \centering
   \includegraphics[width=0.6\textwidth]{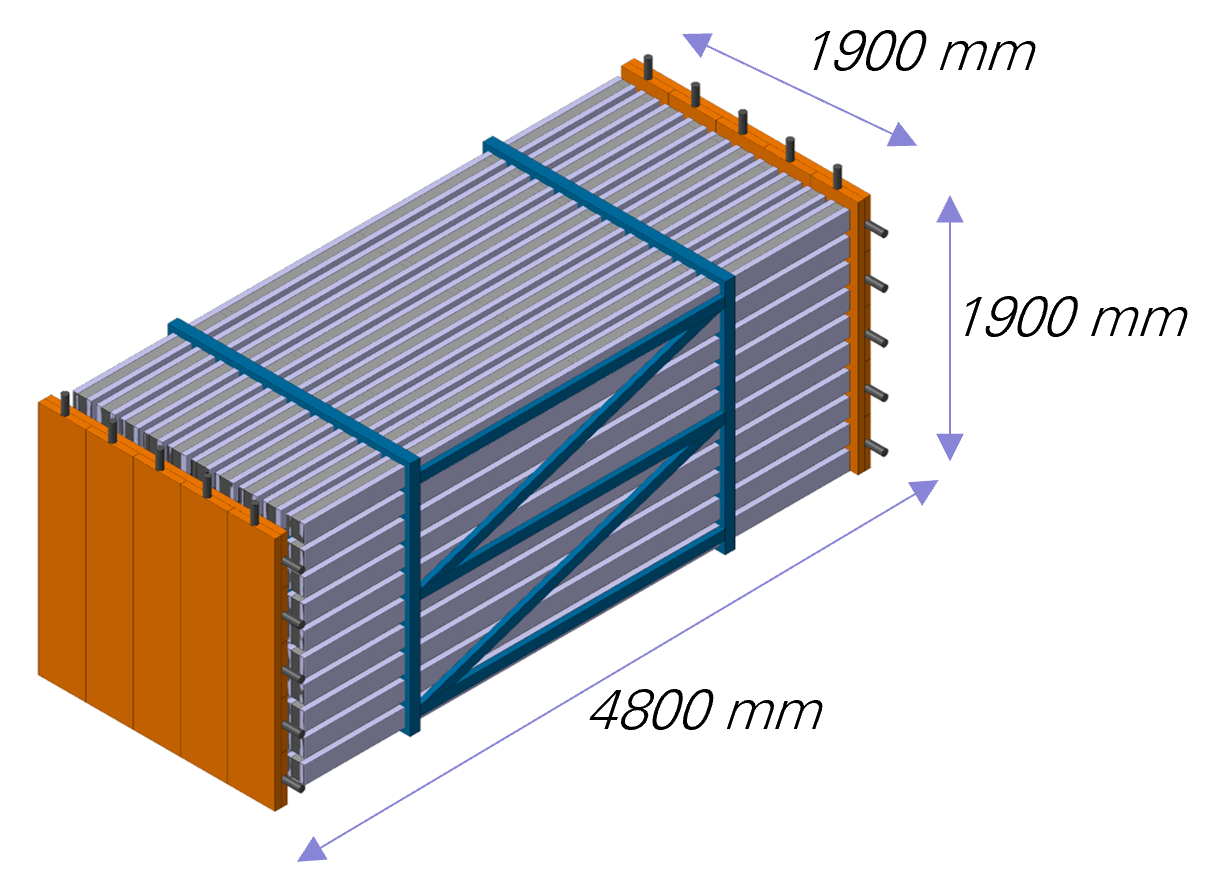}
   \caption{The FORMOSA detector design showing the supermodules (grey), which each hold $2\times2\times4$ scintillator bars, enclosed in a steel frame (blue). Segmented scintillator slabs are shown at the front and back of the detector (orange). For clarity, the top and side scintillator panels are not shown~\cite{Feng_2023}.\label{fig:formosa-bars}}
\end{figure}

The design of the FORMOSA experiment is outlined in Ref.~\cite{Feng_2023}.  The primary feature of the detector is a large quantity of scintillator oriented longitudinally along the direction pointing to the IP. As for the milliQan experiment~\cite{Ball:2016zrp,Ball_2020,Ball_2021}, plastic scintillator is chosen as this provides the best combination of photon yield per unit length, response time, and cost~\cite{adhikary2024}. 
The detector will be comprised of a $1.9~\mathrm{m} \times 1.9~\mathrm{m} \times 4.8~\mathrm{m}$ array of suitable plastic scintillator (e.g., Eljen EJ-200~\cite{Eljen} or Saint-Gobain BC-408~\cite{SG}). The array will be oriented such that the long axis points at the ATLAS IP and will be located on the line of sight (LOS). The array contains four longitudinal ``layers'' arranged to facilitate a 4-fold coincident signal for feebly-interacting particles originating from the ATLAS IP. Each layer in turn contains 400 $5~\mathrm{cm} \times 5~\mathrm{cm} \times 100~\mathrm{cm}$ scintillator ``bars'' in a $20\times20$ array. 
To maximize sensitivity to the smallest charges, each scintillator bar is coupled to a high-gain photomultiplier tube (PMT) capable of efficiently reconstructing the waveform produced by a single photoelectron (PE), such as the Hamamatsu R7725~\cite{Hamamatsu}. 
To reduce random backgrounds, mCP signal candidates will be required to have a quadruple coincidence of hits with $\overline{N}_{\text{PE}} \ge 1$ within a 10-20 ns time window. The PMTs must therefore measure the timing of the scintillator photon pulse with a resolution of $\le5$ ns. 
The bars will be held in place by a steel frame~\cite{adhikary2024}. 
In addition to the scintillator bars, additional components will be installed to reduce or characterize certain types of backgrounds. Scintillator panels on the top and sides of the detector will provide the ability to tag and reject cosmogenic muons and environmental radiation. Finally, segmented scintillator panels will be placed at the front and back of the detector to allow through-going muons to be identified, and their path through the detector tracked. A conceptual design of the FORMOSA detector is shown in Figure~\ref{fig:formosa-bars}. 

The detector readout requires a resolution of $\sim 1\ \mathrm{mV}$ at a sampling rate of $\sim 1\ \mathrm{GS/s}$ in order to efficiently identify and trigger on single photo-electron signals (SPEs). This can be provided by the commercially available CAENV1743 digitizer~\cite{CAENV1743}. Alternatively, a bespoke readout can be deployed at a significantly reduced cost using the DRS4 chip~\cite{DRS4}. Such a design is being utilized by the SUBMET detector at J-PARC~\cite{Kim_2021}.

The detector would be ideally placed in the proposed FPF. This facility would be located $627\ \mathrm{m}$ in front of the ATLAS IP and would provide a dedicated facility for four experiments targeting a wide range of BSM and SM physics~\cite{Feng_2023}. Alternatively, the detector could be located at an existing site near the LHC beamline such as the UJ12 cavern that holds FASER and the current FORMOSA demonstrator (see Sec.~\ref{sec:proto}). Such a site would face increased backgrounds from beam related processes and constraints on the size and position of the detector, however, these challenges may be overcome to provide similar sensitivity to a detector in the FPF. Minor civil engineering work may be required for optimal positioning of the detector along the LOS.

In order to improve the sensitivity to particles with the lowest charges the FORMOSA detector could be augmented by the inclusion of a high performance non-organic scintillator such as CeBr(3)~\cite{BNC}. This material combines a fast response time and low internal radioactivity with a light yield factor $\sim 35$ times that of the same length of EJ-200. Due to the increased costs of such a material this would be a smaller subdetector comprised of an array of up to $4\times4\times4$ bars of CeBr(3). 

\section{Sensitivity to millicharged particles}

\begin{figure}[t]
    \centering
    \includegraphics[width=0.5\textwidth]{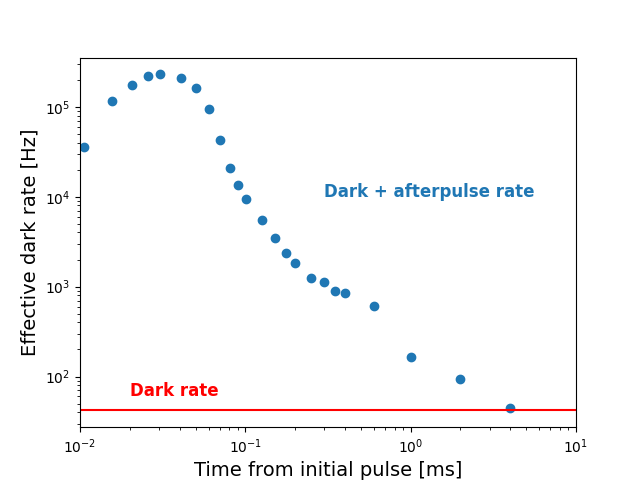}
    \caption{Effective increase in dark rate caused by afterpulses following
    a deposit similar in size to that expected from a through-going muon.
    This is shown to drop below the targeted dark current rate of 1 kHz by $\sim$1 ms.}
    \label{fig:R7725}
\end{figure}
In order to evaluate the sensitivity of the FORMOSA detector, the signal contributions must be reliably simulated. Pair production of millicharged particles of a given mass and charge at the LHC is nearly model independent. Every standard model (SM) process that results in dilepton pairs through a virtual photon would, if kinematically allowed, also produce $\mCPp\mCPm$ pairs with a cross section reduced by a factor of $(Q/e)^2$ and by mass-dependent factors that are well understood. Millicharged particles can also be produced through $Z$ boson couplings that depend on their hypercharge~\cite{Izaguirre:2015eya}. A full consideration of millicharged production mechanisms in the forward region of the LHC through pseudoscalar and vector meson decays has been carried out using Pythia~8~\cite{Skands:2014pea}, while heavier MCP production through Drell-Yan is simulated with MadGraph 5~\cite{Alwall_2011} and Pythia~8~\cite{Skands:2014pea}. For low mass MCPs there will also be contributions from proton-bremsstrahlung, which provides a significant increase in the MCP production for masses below $\sim 1\ \mathrm{GeV}$. The size of the contribution from this process is being determined now.

\begin{figure}[!htb] 
    \centering \includegraphics[width=0.7\columnwidth]{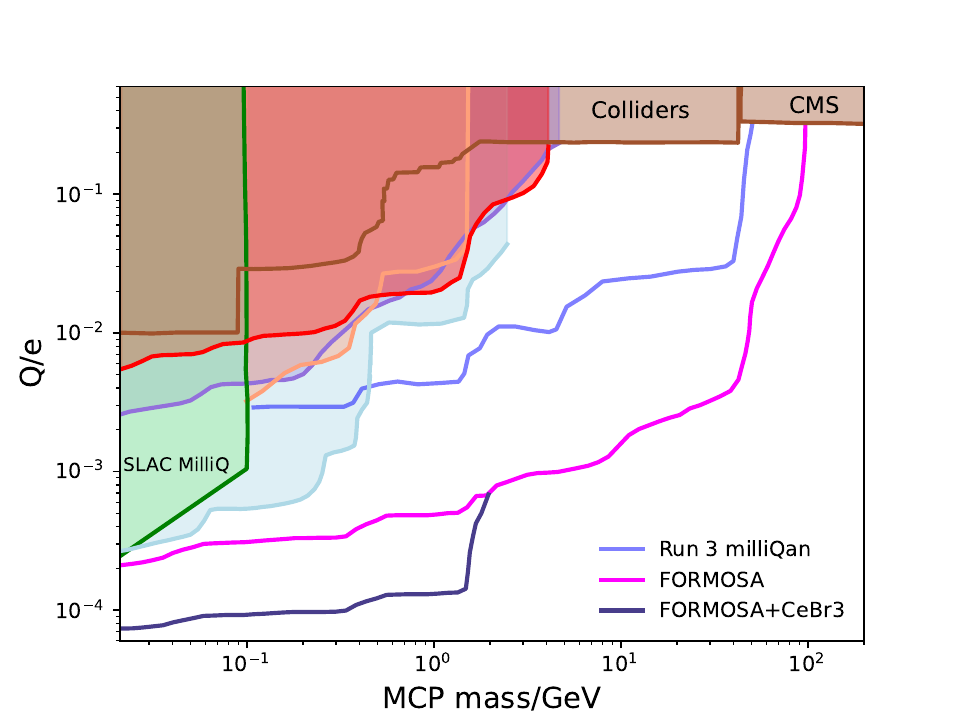}
    \caption{The exclusion limits projected to be achieved by FORMOSA are compared to current bounds~\cite{Plestid_2020,MilliQ,Vogel_2014,essig2013dark,Chatrchyan_2013,Chatrchyan_2013_2,Acciarri_2020, Davidson:2000hf, Badertscher:2006fm,Ball_2020,PhysRevLett.133.071801}. Also shown is the potential sensitivity including a $4\times4$ bar subdetector comprised of CeBr(3). Note that the signal simulation does not proton bremsstrahlung that would increase production for masses below $\sim1$ GeV.\label{fig:limits}}
\end{figure}

The response of the FORMOSA detector to the signal deposits is simulated using \textsc{Geant4}. Backgrounds arise from three main processes: PMT dark rate, showers from cosmogenic muons, and afterpulsing caused by beam muons. Assuming a dark channel rate of 1 kHz (as measured for the R7725 PMTs in-situ), the total background expected over the full HL-LHC running period for the coincidence of four layers within a $10\ \mathrm{ns}$ window is $\sim0.3$ events. The background from showers originated by cosmogenic muons for a similar detector in the milliQan location at the HL-LHC is projected to be $2\times 10^{-5}$ events in Ref.~\cite{Ball_2021} (using a fully calibrated \textsc{Geant4} simulation) after highly efficient signal selection. This background will be a similarly negligible contribution for the FORMOSA detector. Finally, bench tests strongly suggest that contributions from the afterpulses created by beam muons can be made negligible with a small deadtime of $\sim 1\ \mathrm{ms}$ after beam muons are identified passing through the detector (see Fig.~\ref{fig:R7725}), which corresponds to $2.5\%$ deadtime when considering the expected flux of muons traveling through each signal path at the FPF or in the UJ12 location ($\sim 1\ \mathrm{Hz}/\mathrm{cm}^2$).  This is being studied in-situ with the FORMOSA demonstrator as discussed in Sec.~\ref{sec:proto}. The total background is therefore projected to be less than one event over the full HL-LHC data-taking period. In the case that the FORMOSA detector is placed close to the LHC, instead of in the FPF, there will also be backgrounds that originate from the LHC beamline. The impact of these is also being measured by the FORMOSA demonstrator. The expected reach of the FORMOSA detector is shown in Fig.~\ref{fig:limits} as well as the improvement possible with the inclusion of a CeBr(3) subdetector.

\label{sec:sensitivity}
\section{The FORMOSA demonstrator}

\begin{figure}[!htb] 
    \centering \includegraphics[width=0.7\columnwidth]{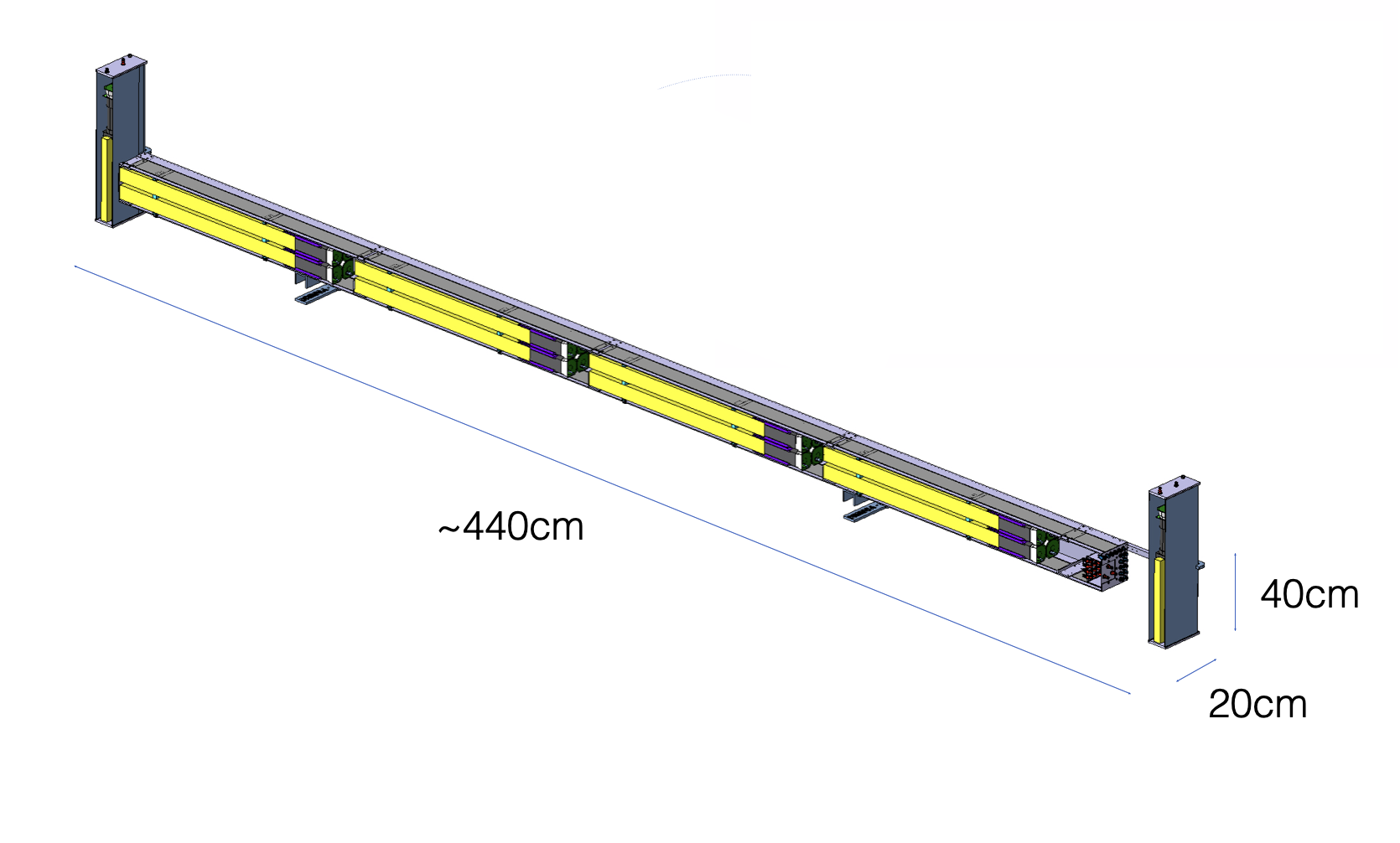}
    \caption{Diagram of the FORMOSA demonstrator showing the bars, and front/back panels.\label{fig:demdiagram}}
\end{figure}

\begin{figure}[!htb] 
     \centering \includegraphics[width=0.7\columnwidth]{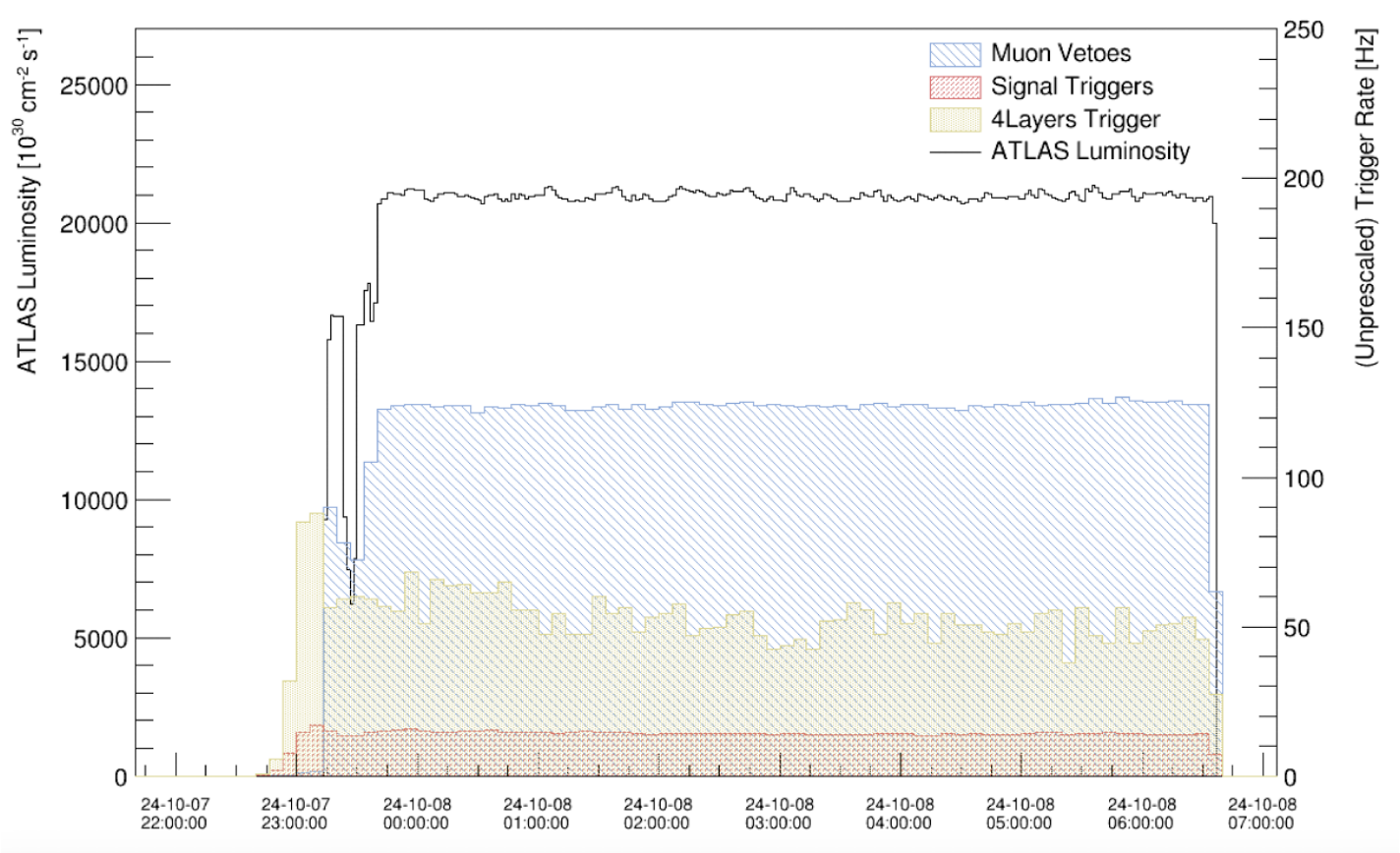}
    \caption{Validation of the DAQ strategy for FORMOSA. The rate of signal triggers is shown to be stable during data taking while contributions from muons and beam related radiation (4 layers trigger) are effectively vetoed.\label{fig:daqvalidation}}
\end{figure}

\begin{figure}[!htb] 
     \centering \includegraphics[width=0.4\columnwidth]{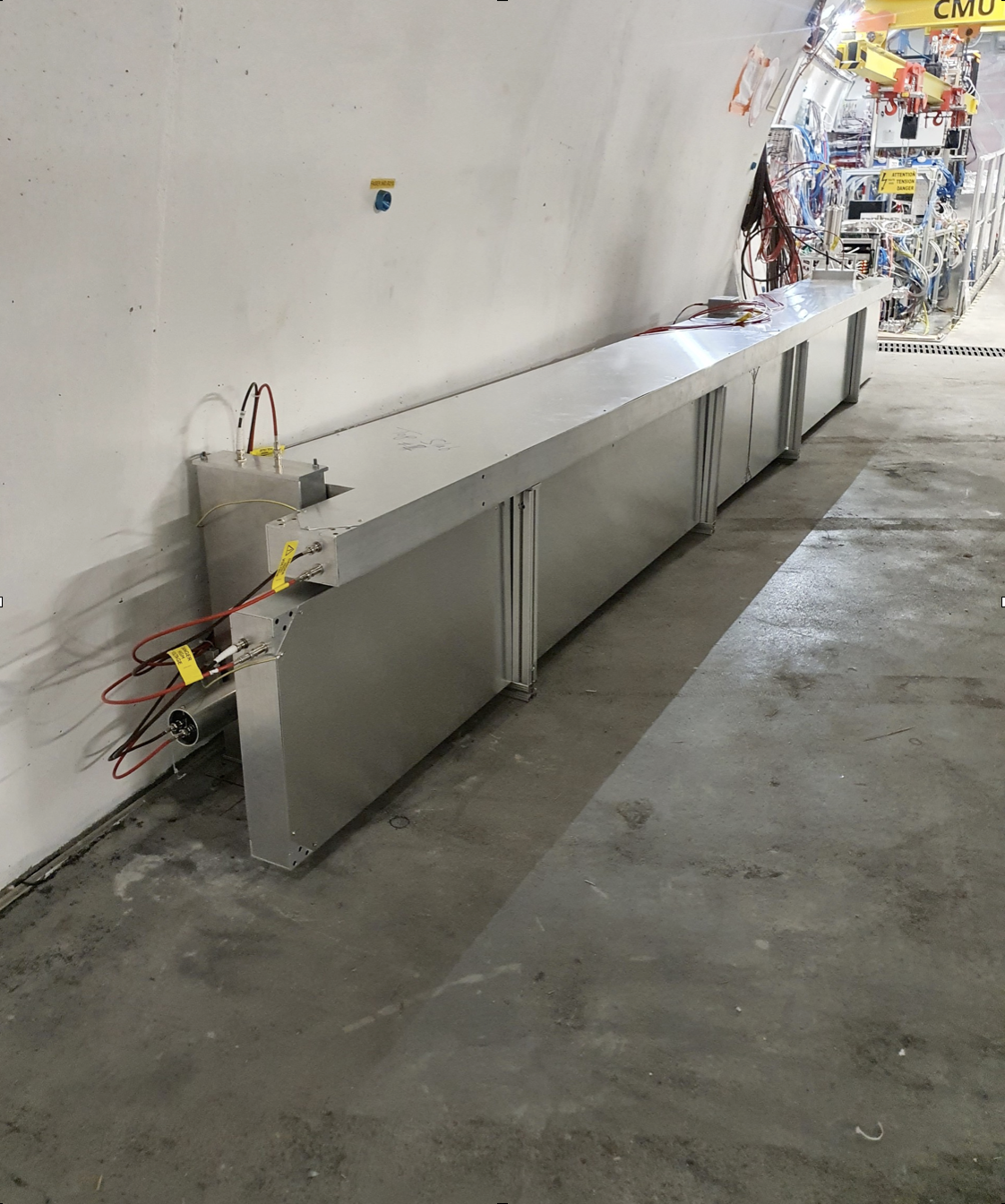}
    \caption{Picture of the upgraded FORMOSA demonstrator in UJ12 showing hermetic veto panels. The CeBr(3) module can be seen at the end.\label{fig:dempicture}}
\end{figure}

\label{sec:proto}
The feasibility of the FORMOSA experiment is being proved through the operation of a demonstrator experiment installed at the LHC in the UJ12 cavern at a distance of $580\ \mathrm{m}$ from the ATLAS IP. The FORMOSA demonstrator is comprised of a $2\times2\times4$ array of $5\ \mathrm{cm}\times 5\ \mathrm{cm}\times 100\ \mathrm{cm}$ scintillator bars. Scintillator panels are placed at the front and back of the bars for efficient tagging of beam muons while a scintillator panel at the side provides tagging and rejection of radiation backgrounds from the beam. The FORMOSA demonstrator was installed in February 2024 and has been commissioned throughout 2024 LHC Run 3 pp collisions. A diagram of the demonstrator is shown in Fig~\ref{fig:demdiagram}. The aims of the demonstrator are to show that signal data may be efficiently collected despite the challenging environment, determine optimal selections to reject backgrounds from beam muon afterpulsing and beam radiation, and to measure any residual backgrounds after these selections. Figure~\ref{fig:daqvalidation} shows the validation of the DAQ strategy with proton-proton collisional data collected during LHC Run 3 in 2024. In February 2025, the side panels were upgraded to be fully hermetic. The upgraded demonstrator is shown in Fig~\ref{fig:dempicture}. The background measurements will be completed with the data collected by the upgraded FORMOSA demonstrator throughout 2025 LHC pp collisions.

In late 2024 a $1\ \mathrm{cm}$ radius $\times~3\ \mathrm{cm}$ long cylindrical module of CeBr(3) scintillator was added to the FORMOSA demonstrator. The performance of this scintillator material is being studied to determine how it may be incorporated into the FORMOSA detector design.

\section{Project planning}
\label{sec:planning}
\begin{description}
    \item[Collaboration] A collaboration of 15 members from 8 institutions, including institutions from the US, Europe and Asia, has been formed to design, construct and operate the FORMOSA experiment. The FORMOSA collaboration is comprised of experts with long-standing experience from related detectors including milliQan~\cite{Ball_2020} and SUBMET~\cite{Kim_2021}. This collaboration has already been able to quickly construct and successfully operate the FORMOSA demonstrator. Monthly meetings and a yearly FORMOSA workshop are held to ensure rapid progress.
    \item[Readiness] The substantial past experience and successful operation of the FORMOSA demonstrator will ensure that the FORMOSA collaboration can rapidly construct a full-scale experiment. The complete conceptual design report can be produced rapidly.
    \item[Budget] The costs for the experiment are well understood from previous efforts. An overall cost of $2.7\rm{MCH}$ is expected to build the detector using CAEN digitizers or $2\rm{MCH}$ using a bespoke readout.
    \item[Sustainability] The dominant energy cost for FORMOSA is in producing the LHC collisions, which will occur in any case. This project therefore helps to optimize the range of physics that can be carried out at the LHC to ensure opportunity for discovery is not missed.
    \item[Technology] The technology outlined for FORMOSA has been proven through past efforts, such as milliQan, and through the FORMOSA demonstrator. Upgrades to the detector design including the use of high performance scintillator will help to study the use of such exotic materials for High Energy physics.
    \item[People and skills] The FORMOSA experiment has already attracted collaborators from around the world. As a small scale experiment it will provide excellent opportunities for leadership to early career researchers. Such efforts are vital to inspire future leaders in HEP.
    \item[Timeliness] The FORMOSA detector can be built within a year of being funded and should be able to take data from the start of LHC Run 4 for optimal physics reach.

\end{description}
\section{Summary}

This contribution outlines the physics case and detector design for the FORMOSA experiment in the far forward region of the ATLAS IP. The FORMOSA detector will help to  ensure that CERN fulfills its mandate to fully exploit the unique opportunities offered by its accelerator complex to complement what can be achieved by the main experiments at the LHC. The FORMOSA experiment at the HL-LHC will provide world-leading sensitivity for millicharged particles that cannot be achieved at any other facility. The experiment is fully costed and a collaboration has been formed comprised of members with significant experience from past millicharge particle experiments. The operation of a demonstrator experiment at the LHC is proving the feasibility of this detector and we will be ready to construct the detector for LHC Run~4. 

\label{sec:summary}
\bibliographystyle{JHEP3}
\bibliography{main}

\providecommand{\noopsort}[1]{}\providecommand{\singleletter}[1]{#1}%

\providecommand{\href}[2]{#2}\begingroup\raggedright\begin{thebibliography}{10}

\bibitem{ArkaniHamed:2008qn}
N.~Arkani-Hamed, D.P.~Finkbeiner, T.R.~Slatyer and N.~Weiner, \emph{{A Theory of Dark Matter}}, \href{https://doi.org/10.1103/PhysRevD.79.015014}{\emph{{Phys. Rev. D}} {\bfseries 79} (2009) 015014} [\href{https://arxiv.org/abs/0810.0713}{{\ttfamily 0810.0713}}].

\bibitem{Pospelov:2008jd}
M.~Pospelov and A.~Ritz, \emph{{Astrophysical Signatures of Secluded Dark Matter}}, \href{https://doi.org/10.1016/j.physletb.2008.12.012}{\emph{{Phys. Lett. B}} {\bfseries 671} (2009) 391} [\href{https://arxiv.org/abs/0810.1502}{{\ttfamily 0810.1502}}].

\bibitem{Battaglieri:2017aum}
M.~Battaglieri, A.~Belloni, A.~Chou, P.~Cushman et~al., \emph{{US Cosmic Visions: New Ideas in Dark Matter 2017: Community Report}},  in \emph{{U.S. Cosmic Visions: New Ideas in Dark Matter}}, Fermilab, Batavia, 2017 [\href{https://arxiv.org/abs/1707.04591}{{\ttfamily 1707.04591}}].

\bibitem{Beacham:2019nyx}
J.~Beacham, C.~Burrage, D.~Curtin, A.~De~Roeck, J.~Evans, J.L.~Feng et~al., \emph{{Physics Beyond Colliders at CERN: Beyond the Standard Model Working Group Report}}, \href{https://doi.org/10.1088/1361-6471/ab4cd2}{\emph{{J. Phys. G}} {\bfseries 47} (2020) 010501} [\href{https://arxiv.org/abs/1901.09966}{{\ttfamily 1901.09966}}].

\bibitem{Strategy:2019vxc}
{European Strategy for Particle Physics Preparatory Group}, \emph{{Physics Briefing Book}},  in \emph{{Physics Briefing Book}}, CERN, Geneva, 2019 [\href{https://arxiv.org/abs/1910.11775}{{\ttfamily 1910.11775}}].

\bibitem{Holdom:1985ag}
B.~Holdom, \emph{{Two U(1)'s and Epsilon Charge Shifts}}, \href{https://doi.org/10.1016/0370-2693(86)91377-8}{\emph{{Phys. Lett. B}} {\bfseries 166} (1986) 196}.

\bibitem{Izaguirre:2015eya}
E.~Izaguirre and I.~Yavin, \emph{{New window to millicharged particles at the LHC}}, \href{https://doi.org/10.1103/PhysRevD.92.035014}{\emph{{Phys. Rev. D}} {\bfseries 92} (2015) 035014} [\href{https://arxiv.org/abs/1506.04760}{{\ttfamily 1506.04760}}].

\bibitem{MilliQ}
A.~Prinz, R.~Baggs, J.~Ballam, S.D.~Ecklund, C.~Fertig, J.A.~Jaros et~al., \emph{{Search for Millicharged Particles at SLAC}}, \href{https://doi.org/10.1103/PhysRevLett.81.1175}{\emph{{Phys. Rev. Lett.}} {\bfseries 81} (1998) 1175} [\href{https://arxiv.org/abs/hep-ex/9804008}{{\ttfamily hep-ex/9804008}}].

\bibitem{essig2013dark}
R.~Essig, J.A.~Jaros, W.~Wester, P.H.~Adrian et~al., \emph{{Working Group Report: New Light Weakly Coupled Particles}},  in \emph{{Community Summer Study 2013}: {Snowmass on the Mississippi}}, Fermilab, Batavia, 10, 2013 [\href{https://arxiv.org/abs/1311.0029}{{\ttfamily 1311.0029}}].

\bibitem{Chatrchyan_2013}
{\scshape {CMS}} collaboration, \emph{{Search for fractionally charged particles in pp collisions at $\sqrt{s}=7$ TeV}}, \href{https://doi.org/10.1103/physrevd.87.092008}{\emph{{Phys. Rev. D}} {\bfseries 87} (2013) } [\href{https://arxiv.org/abs/1210.2311}{{\ttfamily 1210.2311}}].

\bibitem{Chatrchyan_2013_2}
{\scshape {CMS}} collaboration, \emph{{Searches for long-lived charged particles in pp collisions at $ \sqrt{s} $ = 7 and 8 TeV}}, \href{https://doi.org/{10.1007/jhep07(2013)122}}{\emph{{J. High Energy Phys.}} {\bfseries 07} ({2013}) 122} [\href{https://arxiv.org/abs/1305.0491}{{\ttfamily 1305.0491}}].

\bibitem{Acciarri_2020}
R.~Acciarri, C.~Adams, J.~Asaadi, B.~Baller, T.~Bolton, C.~Bromberg et~al., \emph{{Improved Limits on Millicharged Particles Using the ArgoNeuT Experiment at Fermilab}}, \href{https://doi.org/10.1103/physrevlett.124.131801}{\emph{{Phys. Rev. Lett.}} {\bfseries 124} (2020) } [\href{https://arxiv.org/abs/1911.07996}{{\ttfamily 1911.07996}}].

\bibitem{Davidson:2000hf}
S.~Davidson, S.~Hannestad and G.~Raffelt, \emph{{Updated bounds on milli-charged particles}}, \href{https://doi.org/10.1088/1126-6708/2000/05/003}{\emph{{J. High Energy Phys.}} {\bfseries 05} (2000) 003} [\href{https://arxiv.org/abs/hep-ph/0001179}{{\ttfamily hep-ph/0001179}}].

\bibitem{Badertscher:2006fm}
A.~Badertscher, P.~Crivelli, W.~Fetscher, U.~Gendotti, S.~Gninenko, V.~Postoev et~al., \emph{{An Improved Limit on Invisible Decays of Positronium}}, \href{https://doi.org/10.1103/PhysRevD.75.032004}{\emph{{Phys. Rev. D}} {\bfseries 75} (2007) 032004} [\href{https://arxiv.org/abs/hep-ex/0609059}{{\ttfamily hep-ex/0609059}}].

\bibitem{Magill_2019}
G.~Magill, R.~Plestid, M.~Pospelov and Y.-D.~Tsai, \emph{Millicharged particles in neutrino experiments}, \href{https://doi.org/10.1103/physrevlett.122.071801}{\emph{Phys. Rev. Lett.} {\bfseries 122} (2019) }.

\bibitem{Ball:2016zrp}
A.~Ball, J.~Brooke, C.~Campagnari, A.D.~Roeck et~al., \emph{{A Letter of Intent to Install a Milli-Charged Particle Detector at LHC P5}},  2016.

\bibitem{Ball_2020}
A.~Ball, G.~Beauregard, J.~Brooke, C.~Campagnari, M.~Carrigan, M.~Citron et~al., \emph{{Search for millicharged particles in proton-proton collisions at $\sqrt{s}=13$ TeV}}, \href{https://doi.org/10.1103/physrevd.102.032002}{\emph{Physical Review D} {\bfseries 102} (2020) }.

\bibitem{Ball_2021}
A.~Ball, J.~Brooke, C.~Campagnari, M.~Carrigan, M.~Citron, A.~De~Roeck et~al., \emph{Sensitivity to millicharged particles in future proton-proton collisions at the lhc with the milliqan detector}, \href{https://doi.org/10.1103/physrevd.104.032002}{\emph{Physical Review D} {\bfseries 104} (2021) }.

\bibitem{Tsai:2020ForMINI}
S.~Foroughi-Abari, F.~Kling and Y.-D.~Tsai, \emph{Looking forward to millicharged dark sectors at the lhc}, \href{https://doi.org/10.1103/physrevd.104.035014}{\emph{Physical Review D} {\bfseries 104} (2021) }.

\bibitem{Feng_2023}
J.L.~Feng, F.~Kling, M.H.~Reno, J.~Rojo, D.~Soldin, L.A.~Anchordoqui et~al., \emph{The forward physics facility at the high-luminosity lhc}, \href{https://doi.org/10.1088/1361-6471/ac865e}{\emph{Journal of Physics G: Nuclear and Particle Physics} {\bfseries 50} (2023) 030501}.

\bibitem{Feng_2024}
J.L.~Feng, J.~Li, X.~Liao, J.~Ni and J.~Pei, \emph{Discovering quirks through timing at faser and future forward experiments at the lhc}, \href{https://doi.org/10.1007/jhep06(2024)197}{\emph{Journal of High Energy Physics} {\bfseries 2024} (2024) }.

\bibitem{Asai:2905898}
{\scshape P5} collaboration, S.~Asai, A.~Ballarino, T.~Bose, K.~Cranmer, F.-Y.~Cyr-Racine, S.~Demers et~al., \emph{{Exploring the Quantum Universe: Pathways to Innovation and Discovery in Particle Physics. Pathways to Innovation and Discovery in Particle Physics. Report of the 2023 Particle Physics Project Prioritization Panel (P5)}},  2023.
\newblock 10.2172/2368847.

\bibitem{adhikary2024}
J.~Adhikary, L.A.~Anchordoqui, A.~Ariga, T.~Ariga, A.J.~Barr, B.~Batell et~al., \emph{Science and project planning for the forward physics facility in preparation for the 2024-2026 european particle physics strategy update},  2024.

\bibitem{Eljen}
``{Eljen Technology}.'' \url{https://eljentechnology.com/products/plastic-scintillators/ej-200-ej-204-ej-208-ej-212}, Accessed: \today.

\bibitem{SG}
``{Saint-Gobain}.'' \url{https://www.crystals.saint-gobain.com/products/bc-408-bc-412-bc-416}, Accessed: \today.

\bibitem{Hamamatsu}
``{Hamamatsu Photonics K.K.}.'' \url{https://www.hamamatsu.com/}, Accessed: \today.

\bibitem{CAENV1743}
``{CAEN S.p.A}.'' \url{https://www.caen.it/products/v1743/}, Accessed: \today.

\bibitem{DRS4}
``{DRS4}.'' \url{https://www.psi.ch/en/drs}, Accessed: \today.

\bibitem{Kim_2021}
J.H.~Kim, I.S.~Hwang and J.H.~Yoo, \emph{Search for sub-millicharged particles at j-parc}, \href{https://doi.org/10.1007/jhep05(2021)031}{\emph{Journal of High Energy Physics} {\bfseries 2021} (2021) }.

\bibitem{BNC}
``{Berkeley Nucleonics}.'' \url{https://www.berkeleynucleonics.com/cerium-bromide}, Accessed: \today.

\bibitem{Skands:2014pea}
P.~Skands, S.~Carrazza and J.~Rojo, \emph{{Tuning PYTHIA 8.1: the Monash 2013 Tune}}, \href{https://doi.org/10.1140/epjc/s10052-014-3024-y}{\emph{{Eur. Phys. J. C}} {\bfseries 74} (2014) 3024} [\href{https://arxiv.org/abs/1404.5630}{{\ttfamily 1404.5630}}].

\bibitem{Alwall_2011}
J.~Alwall, M.~Herquet, F.~Maltoni, O.~Mattelaer and T.~Stelzer, \emph{Madgraph 5: going beyond}, \href{https://doi.org/10.1007/jhep06(2011)128}{\emph{Journal of High Energy Physics} {\bfseries 2011} (2011) }.

\bibitem{Plestid_2020}
R.~Plestid, V.~Takhistov, Y.-D.~Tsai, T.~Bringmann, A.~Kusenko and M.~Pospelov, \emph{Constraints on millicharged particles from cosmic-ray production}, \href{https://doi.org/10.1103/physrevd.102.115032}{\emph{Physical Review D} {\bfseries 102} (2020) }.

\bibitem{Vogel_2014}
H.~Vogel and J.~Redondo, \emph{Dark radiation constraints on minicharged particles in models with a hidden photon}, \href{https://doi.org/{10.1088/1475-7516/2014/02/029}}{\emph{{J. Cosmol. Astropart. Phys.}} {\bfseries 02} ({2014}) 029} [\href{https://arxiv.org/abs/1311.2600}{{\ttfamily 1311.2600}}].

\bibitem{PhysRevLett.133.071801}
{\scshape SENSEI Collaboration} collaboration, \emph{Search by the sensei experiment for millicharged particles produced in the numi beam}, \href{https://doi.org/10.1103/PhysRevLett.133.071801}{\emph{Phys. Rev. Lett.} {\bfseries 133} (2024) 071801}.

\end{thebibliography}\endgroup

\end{document}